\documentclass[12pt,aps,prb,preprint]{revtex4-1}
\usepackage{epsfig} 
\usepackage{amsmath,amssymb}
\usepackage{graphicx}
\usepackage{caption}
\usepackage{subcaption}

\newcommand{\be}{\begin{equation}}
\newcommand{\ee}{\end{equation}} 
\newcommand{\bq}{\begin{eqnarray}} 
\newcommand{\eq}{\end{eqnarray}}

\begin{document}

\begin{center} {\bf Frictionless Racquetball?} \end{center}
\author{Robert Brashear, Kevin Kim, Carolina Santos, Mikhail Kagan} 
\maketitle

\section*{Introduction} 
When a ball hits a surface, does the angle of reflection always equal the angle of incidence? Not at all! Depending on the interplay of the ball's spin and speed and the force of friction, the ball's behavior after the bounce may differ dramatically. Surely some experienced racquetball, ping-pong, pool, tennis and players alike take advantage of this fact. A physics teacher, in turn, can take advantage of the fact that by observing the bounce of a ball her students can determine the coefficient of friction between the ball and the floor. All it takes is a typical video (smartphone) camera and some standard software \cite{Logger}.

Indeed during a bounce, the force of friction changes the tangential velocity of the ball, while the normal force affects its perpendicular component. Hence, measuring the ball's initial and final velocity, one can compute the coefficient of friction. 

In fact, this would be a routine exercise, if not for the following intriguing effect. When trying to repeat the same calculations for a few consecutive bounces of the same ball, friction seems to dissapear after the first bounce!

The paper is organized as follows. We start by considering a single bounce of a racquetball and compute the coefficent of friction between the ball and the floor (Sec. \ref{Sec:SimpleBounce}). In Sec. \ref{Sec:MultiBounce} we try to reproduce the result by doing identical calculations for the second, third and so on bounces of the same ball and discover the ``absence'' of friction force. In Sec. \ref{Sec:Hypothesis} we hypothesize why this could occur and run two testing experiments (Sec. \ref{Sec:Testing}) to see if the hypothesis is supported. 
These new results suggest that the actual value of the  coefficient of friction obtained in Sec. \ref{Sec:SimpleBounce} should be reconsidered, which we do in Sec. \ref{Sec:HS_COF}. We conclude, in Sec. \ref{Sec:Discussion}, with a discussion of some subtleties and implicit assumtions, as well as how the topic of the paper can be turned into an in-class activity.

\section{A simple bounce}\label{Sec:SimpleBounce}
\begin{figure}[h] 
\centerline{\includegraphics[width=6cm, keepaspectratio]{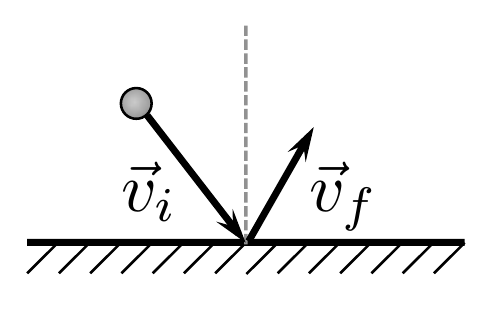}} \caption{Bouncing ball. In general, neither $x$- nor $y$-component of the ball's initial velocity equals those of the final velocity.\label{Fig:SimpleBounce}} 
\end{figure}
As mentioned above, the coefficient of friction between a racquetball and the floor can be determined with the help of the experiment shown in Fig. \ref{Fig:SimpleBounce}. Let $\vec v_i=(v_{ix},v_{iy})$ and $\vec v_f=(v_{fx},v_{fy})$ be the initial and final ball's velocity respectively. These quantities can be found using video analysis, as we explain below. During the bounce, the horizontal and vertical velocities/momenta of the ball are changing due to the force of friction ($f$) and normal force ($N$) respectively. Since the two forces are acting over the same period of time, the coefficient of friction can be found as
\be\label{CoF}
\mu = \frac{f}{N}=\frac{\Delta p_x/\Delta t}{\Delta p_y/\Delta t}=\frac{\Delta p_x}{\Delta p_y}=\frac{v_{xf}-v_{xi}}{v_{yf}-v_{yi}},
\ee
using  the initial and final $x$- and $y$-components of the ball's velocity. Note that in general the final angle is {\em not} equal to the initial angle.  

\begin{figure}
        \centering
        \begin{subfigure}[b]{0.4\textwidth}
                \centering
                \frame{\includegraphics[width=\textwidth]{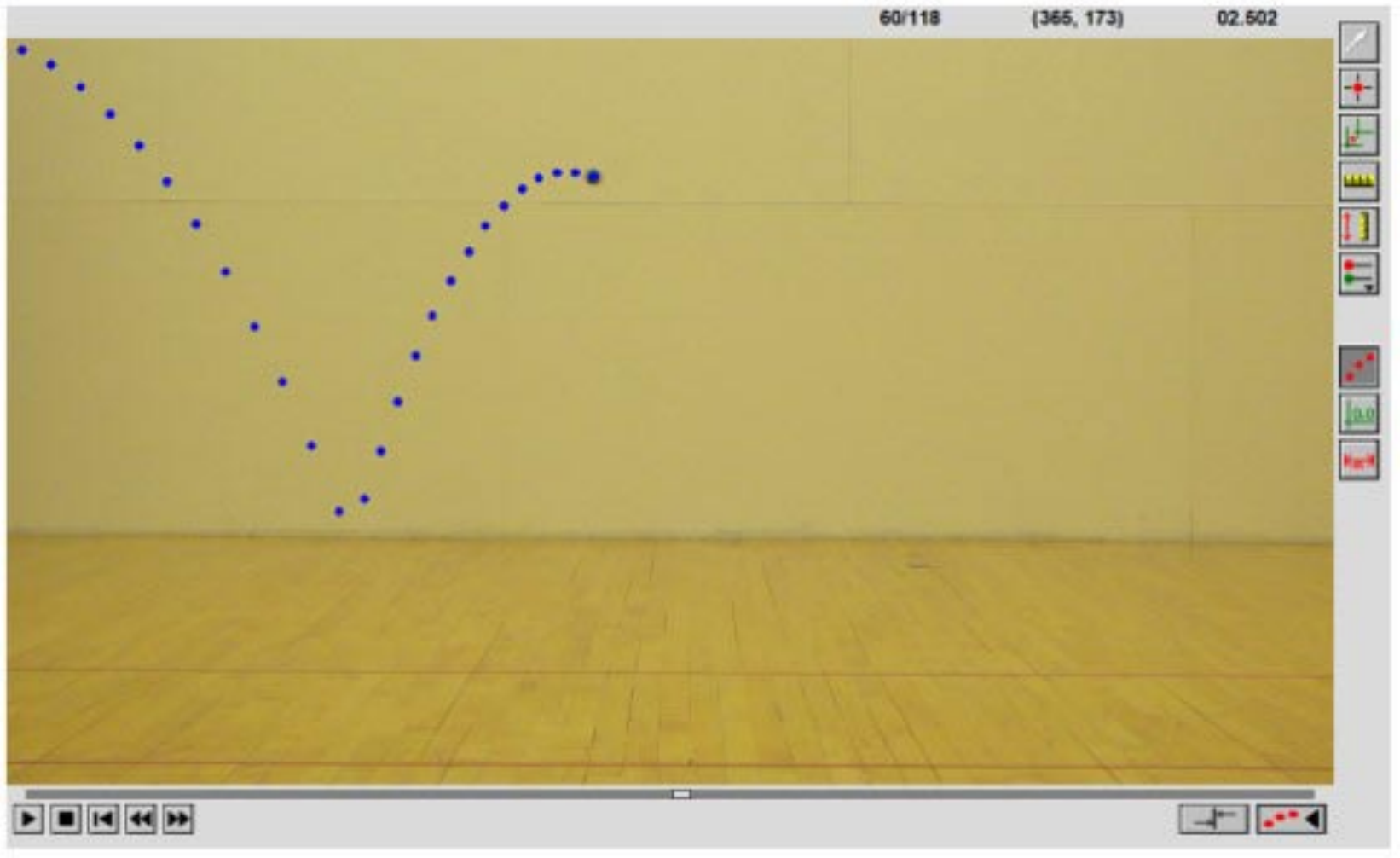}}
                \caption{Ball's trajectory}
                \label{Fig:SimpleBounceTraj}
        \end{subfigure}%
		\qquad
        ~ 
        \begin{subfigure}[b]{0.4\textwidth}
                \centering
                \frame{\includegraphics[width=\textwidth]{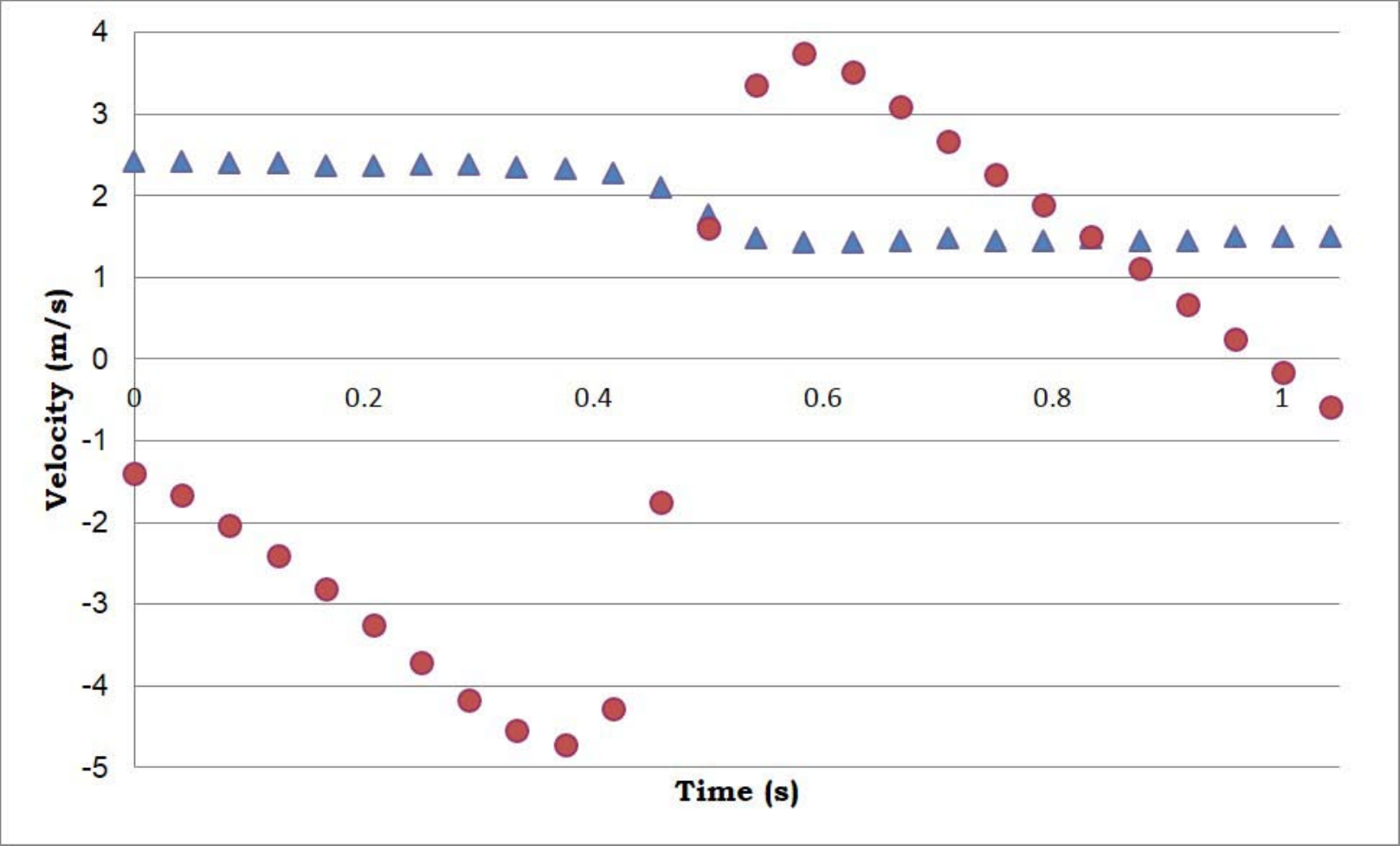}}
                \caption{Velocity graphs}
                \label{Fig:SimpleBounceVelocity}
        \end{subfigure}
        \caption{Video data for the bounce in Fig. \ref{Fig:SimpleBounce}. The time interval between each two consecutive data points is 1/30 s.}\label{Fig:Speed_Spin}
\end{figure}

We have recorded such a bounce using a video camera. The results are displayed below. In Fig. \ref{Fig:SimpleBounceTraj} you can see the (tapped) trajectory of the racquetball, and in Fig. \ref{Fig:SimpleBounceVelocity} we present the corresponding graphs of the horizontal and vertical velocity. Substituting the data into Eq. (\ref{CoF}), we obtain the coefficient of friction between the ball and the floor to be 
\[
\mu=0.106.
\]

It is certainly pertinent to repeat the experiment and to take several measurements to achieve a better accuracy. On the other hand, what can be simpler than considering consecutive bounces of the same toss? We can repeat the same analysis for the second, third etc. bounce and check if we get a consistent value for the coefficient of friction. This is done in the following section.

\section{Mulitple bounces}\label{Sec:MultiBounce}
\begin{figure}[h]
        \centering
        \begin{subfigure}[b]{0.4\textwidth}
                \centering
                \frame{\includegraphics[width=\textwidth]{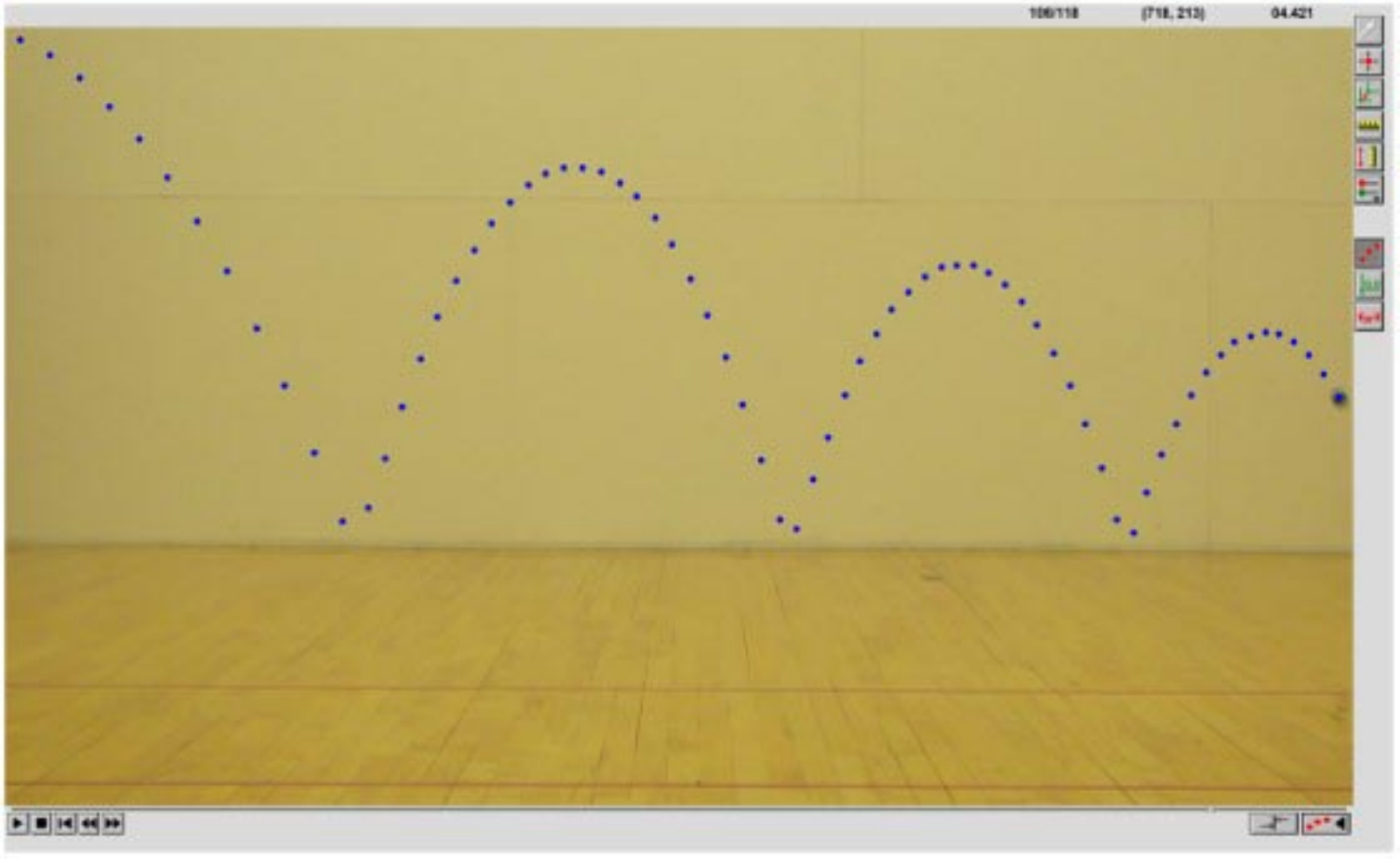}}
                \caption{Ball's trajectory}
                \label{Fig:MultiBounceTraj}
        \end{subfigure}%
		\qquad
        ~ 
        \begin{subfigure}[b]{0.4\textwidth}
                \centering
                \frame{\includegraphics[width=\textwidth]{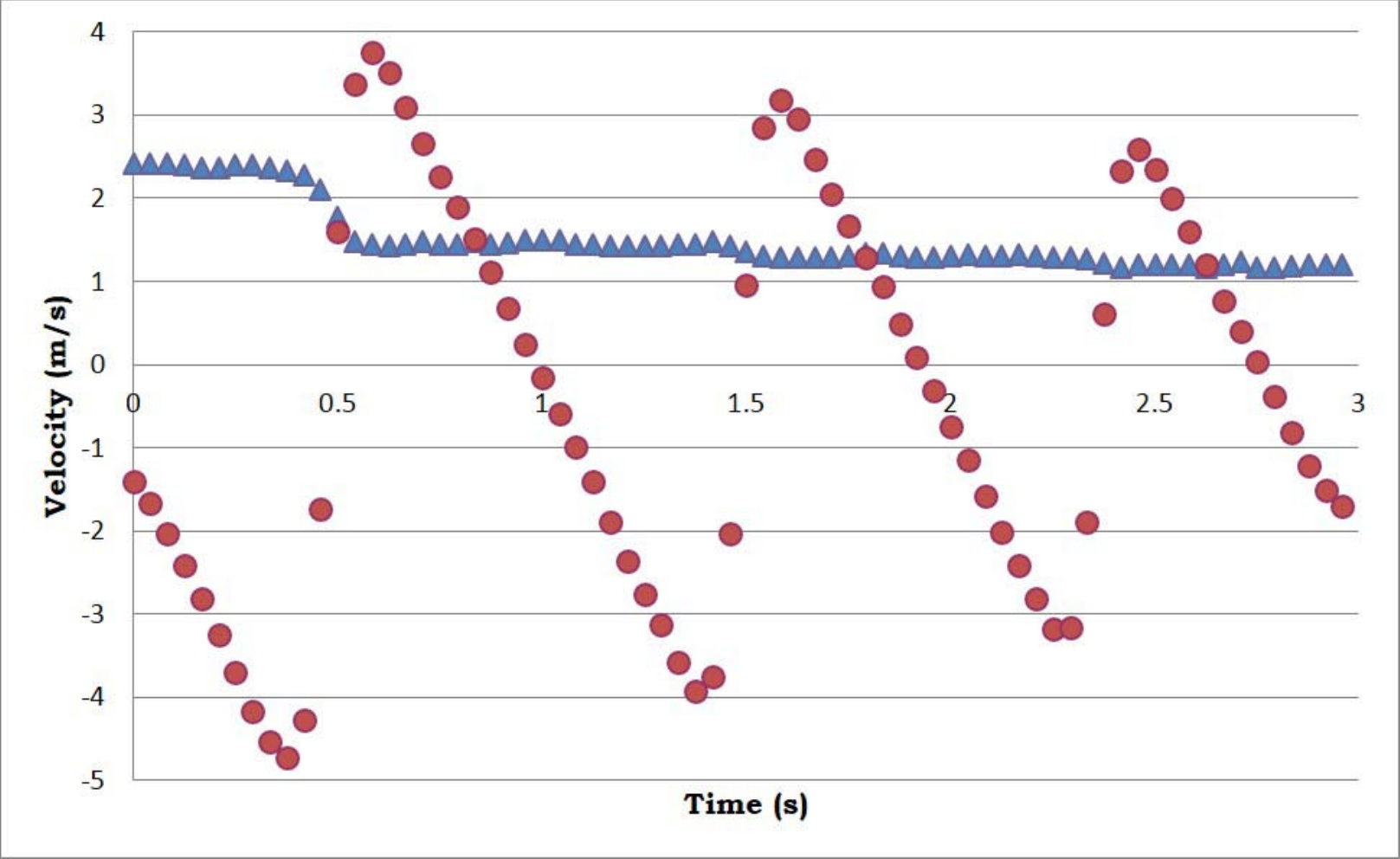}}
                \caption{Velocity graphs}
                \label{Fig:MultiBounceVelocity}
        \end{subfigure}
        \caption{Video data for multiple bounces. The time interval between each two consecutive data points is 1/30 s.}\label{Fig:Speed_Spin}
\end{figure}

Fig. \ref{Fig:MultiBounceTraj} depicts the (dotted) trajectory of a racquetball undergoing several consecutive bounces. The corresponding velocity graphs are displayed in Fig. \ref{Fig:MultiBounceVelocity}. Judging by the jumps in the $y$-velocity (as well as by the trajectory in Fig. \ref{Fig:MultiBounceTraj}), there were three bounces. While the first bounce looks pretty much the same as before, there is something surprising about the next recorded bounces: there is virtually no change in the $x$-velocity after the first bounce. Did someone turn friction off?!

\section{What happens to friction in the second bounce and on? A hypothesis.}\label{Sec:Hypothesis}
In order to understand what happens at the second bounce and later we should recall that the force of friction is acting at the surface of contact, essentially the bottom point of the racquetball. Thus the force of friction is determined by the bottom point's velocity, which in general can take on an arbitrary value: positive, negative or zero with respect to the chosen $x$-axis. \\

\begin{figure}
        \centering
        \begin{subfigure}[b]{0.32\textwidth}
                \centering
                \includegraphics[width=\textwidth]{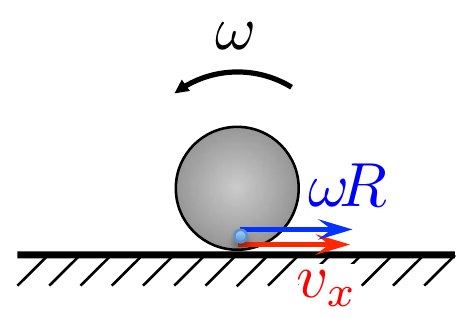}
                \caption{Backspin. The bottom point slides forward.}
                \label{Fig:Speed_Spin_Back}
        \end{subfigure}%
        ~ 
        \begin{subfigure}[b]{0.32\textwidth}
                \centering
                \includegraphics[width=\textwidth]{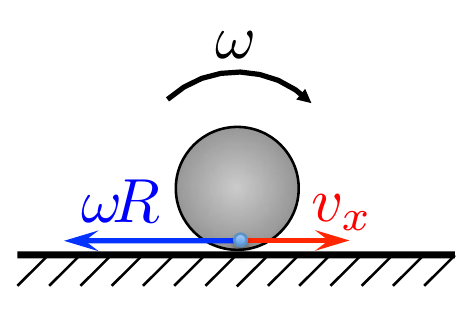}
                \caption{Fast topspin. The bottom point slides backward.}
                \label{Fig:Speed_Spin_Top}
        \end{subfigure}
        ~ 
        \begin{subfigure}[b]{0.32\textwidth}
                \centering
                \includegraphics[width=\textwidth]{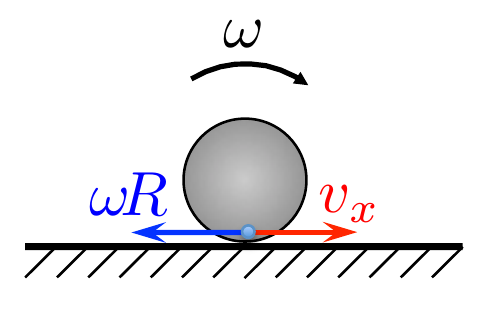}
                \caption{Balanced topspin. The bottom point does not slide.}
                \label{Fig:Speed_Spin_Smooth}
        \end{subfigure}
        \caption{Types of contact}\label{Fig:Speed_Spin}
\end{figure}

For example, any ping-pong player would know that a back-spinning ball or a ball with a small top spin would slow-down after hitting the table. This signals that friction is opposing the ball's horizontal translational velocity, which in turn implies that the bottom point is sliding {\em forward} (see Fig. \ref{Fig:Speed_Spin_Back}). On the other hand, giving enough top spin to the ball would ensure that the bottom point would slide {\em backward} (Fig. \ref{Fig:Speed_Spin_Top}). This would cause friction go with the ball's translational velocity thus speeding it up. It is also clear that there exists a special case of ``balanced'' speed and spin, such that the bottom point is {\em not sliding} across the surface (see Fig. \ref{Fig:Speed_Spin_Smooth}). This happens if 
\be\label{Smooth}
v_x = \omega R,
\ee
where $\omega$ is the angular speed of the ball and $R$ is its radius. Thus condition (\ref{Smooth}) would ``turn friction off''.
\\

In the example considered in the previous section, the ball is not spinning before the first bounce, $\omega_{1i}=0$, $v_{1xi}>0$. Thus the bottom point is sliding forward and friction is acting in the opposite direction. In fact, friction does two things during the first bounce: i) slowing down the ball's translational movement and ii) increasing its clockwise spin. So it is possible that the ``frictionless'' condition (\ref{Smooth}) condition is met before or at the end of the bounce. Even if former is the case, the friction will cease and neither horizontal velocity nor angular velocity will change during the rest of the bounce. Therefore, at the end of the first bounce we shall have
\[
v_{1xf} = \omega_{1f} R.
\]
Now neglecting air resistance, neither of the two variables will be affected between the bounces. Thus  the second bounce will start with 
\[
v_{2if} = \omega_{2i} R,
\]
and friction will not turn on. Obviously, the same will be true for any consecuitive bounce. \\

In summary, we propose the following hypothesis: {\em i) If the the horizontal and angular velocity satisfy (\ref{Smooth}) the ball will not experience (kinetic) friction; ii) The force of friction acting on the ball during the first bounce is large enough to bring the horizontal and spinning speed to balance (condition (\ref{Smooth})) within the duration of the bounce. Once such condition is met, it remains true at later times and friction does not ``turn on'' in the following bounces.}

\section{How can we test the hypothesis? A new experiment}\label{Sec:Testing}
Let us first verify that arranging the right mix of speed and spin will result in a frictionless bounce. Consider a ball rolling smoothly (without slipping) across the table in Fig. \ref{Fig:TableRoll}. If this ball falls off the table, it should have horizontal and angular velocities satisfying (\ref{Smooth}) when hitting the floor. It should, therefore, undergo a frictionless bounce, and we would not see any change in horizontal speed even during the first bounce.

\begin{figure}[h!] 
\centerline{\includegraphics[width=6cm, keepaspectratio]{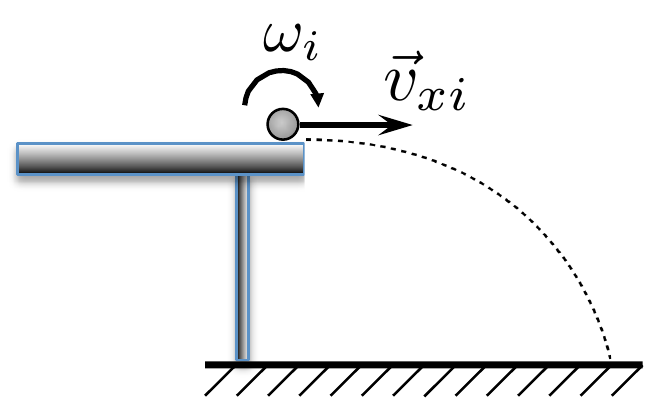}} \caption{Ball rolling off the table. Before leaving the table, the ball rolls without slipping.\label{Fig:TableRoll}} 
\end{figure}

\begin{figure}
        \centering
        \begin{subfigure}[b]{0.4\textwidth}
                \centering
                \frame{\includegraphics[width=\textwidth]{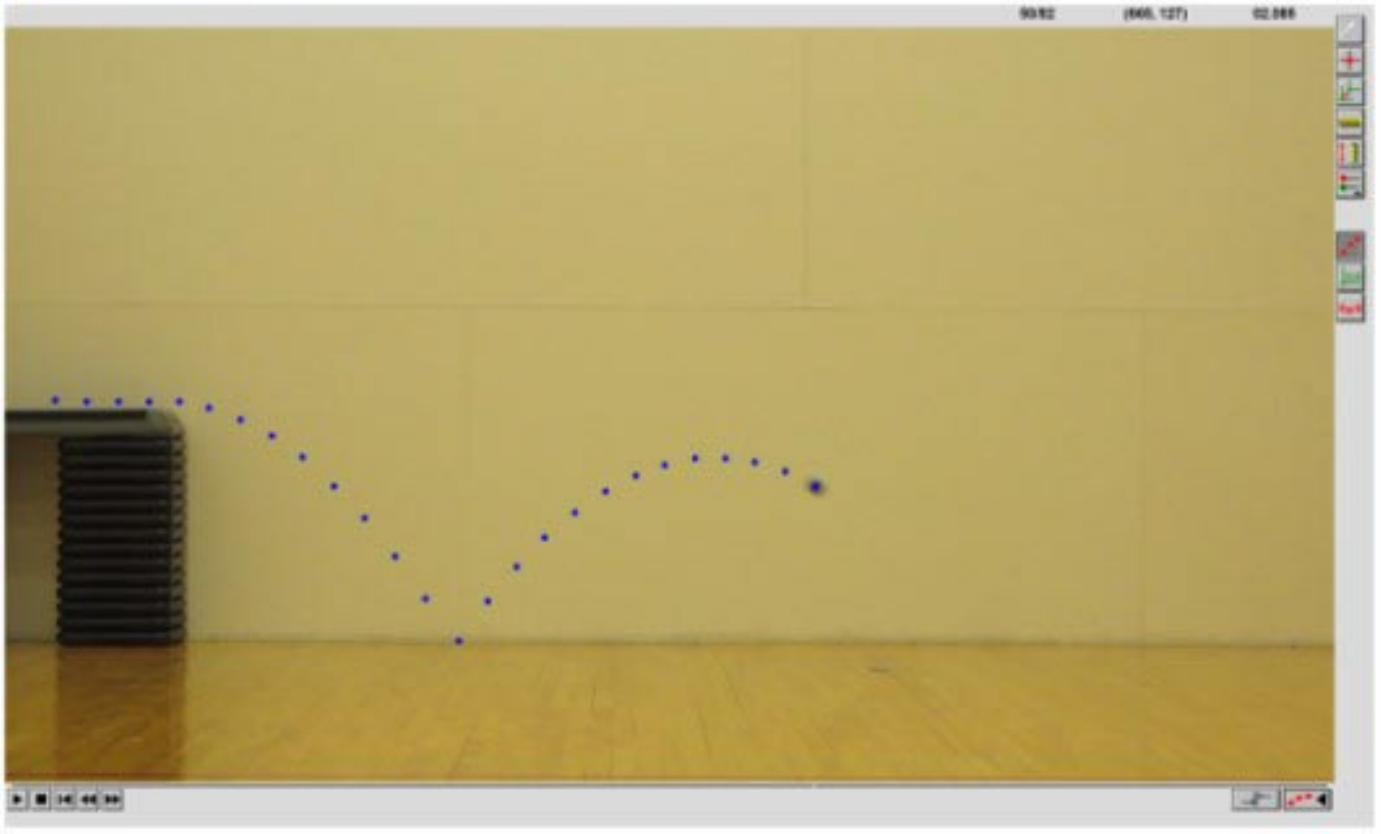}}
                \caption{Ball's trajectory}
                \label{Fig:TableRollTraj}
        \end{subfigure}%
		\qquad
        ~ 
        \begin{subfigure}[b]{0.4\textwidth}
                \centering
                \frame{\includegraphics[width=\textwidth]{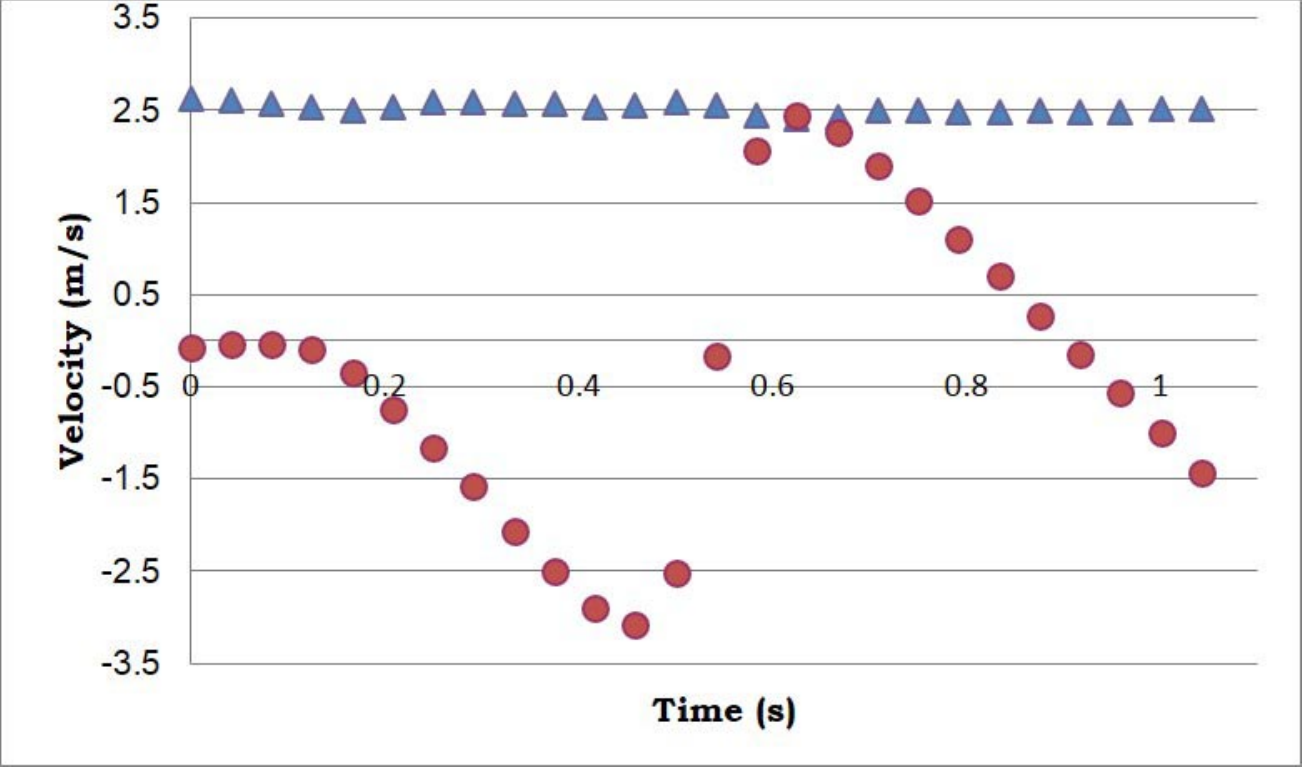}}
                \caption{Velocity graphs}
                \label{Fig:TableRollVelocity}
        \end{subfigure}
        \caption{Video data for the ball rolling off the table. The time interval between each two consecutive data points is 1/30 s.}\label{Fig:TableRoll}
\end{figure}

We have conducted such an experiment. Fig. \ref{Fig:TableRollTraj} depicts the trajectory of the ball and Fig. \ref{Fig:TableRollVelocity} displays the horizontal and vertical velocities of the ball. Based on the vertical velocity graph, we can clearly see that the bounce occurred at around $t=0.5s$. Importantly, there is no substantial change in the horizontal velocity. Thus the first part of the hypothesis is verified: if condition (\ref{Smooth}) is met, the bounce occurs without friction.

We are now to verify that the force of friction is sufficient to stop the ball's slipping within the duration of the first bounce. What kind of experiment would we need to conduct in order to show that? Below we present two options.

The force of friction is proportional to the ball's horizontal acceleration. Thus, the most direct way of checking whether friction turns off during the first bounce would be extracting the acceleration data from the first bounce. Unfortunately, the bounce happens very quickly (5-10 ms), so a typical video camera, making 30 frames per second (fps), would only yield a few data points (three in our case). In order to make the above method work, one would need a high-speed camera. This is done in the next section.  Suppose, however, that we are stuck with the equipment at hand. Is there any way of checking the hypothesis?

Imagine that we have a means of measuring the ball's horizontal velocity and angular speed {\em after} the first bounce. If we find that they satisfy (\ref{Smooth}), given the fact that they did not before the first bounce,  that would support the hypothesis. While obtaining the horizontal velocity is straighforward, measuring angular speed is not easy. That would require a finely tuned adjustment such that the plane of the ball's trajectory coincides with its plane of rotation and remains perpendicular to the camera's line of sight. Furthermore there should be a clear mark that would stay on the {\em rim} of the ball throughout a substantial time period after the first bounce. Nevertheless, these difficulties can be overcome and such an experiment can be realized with a reasonable accuracy if instead of a rubber ball we use a wheel.

\begin{figure}
        \centering
        \begin{subfigure}[b]{0.45\textwidth}
                \centering
                \frame{\includegraphics[width=\textwidth]{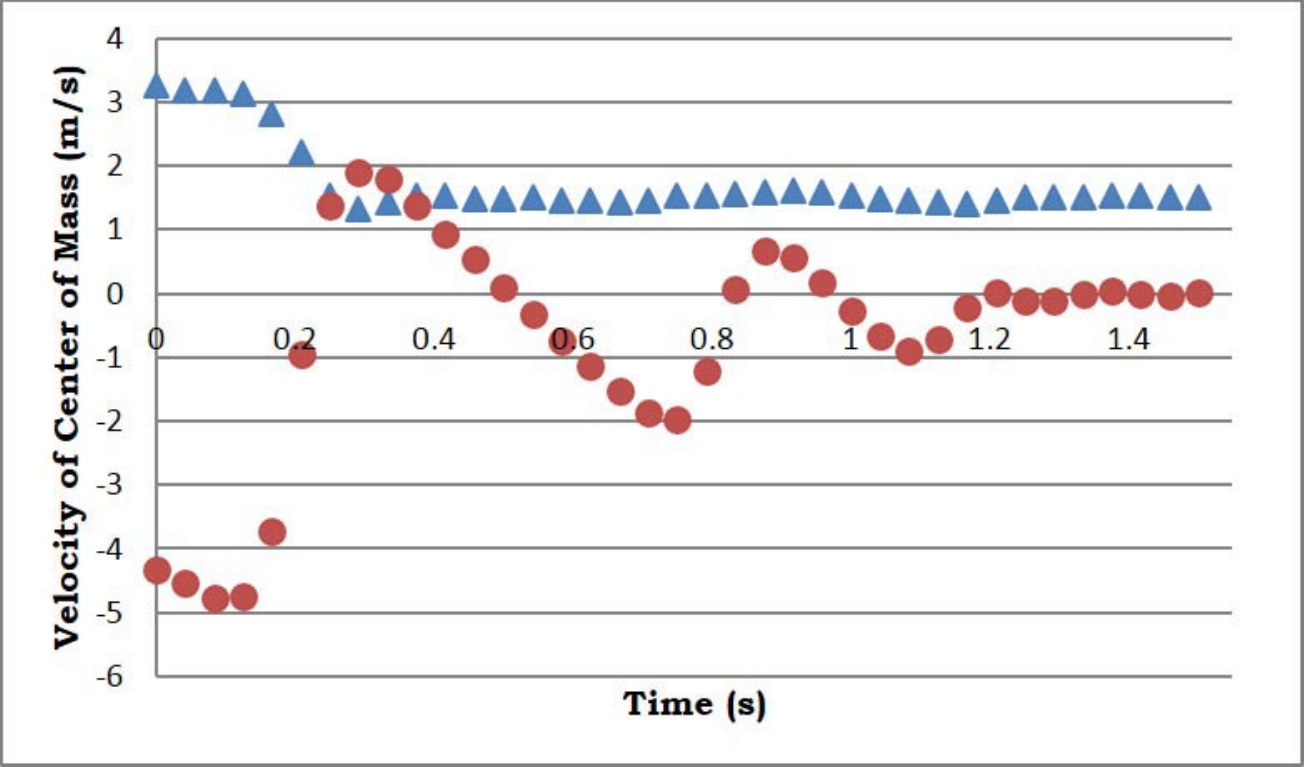}}
                \caption{{\footnotesize Wheel's velocity graphs}}
                \label{Fig:WheelVelocity}
        \end{subfigure}%
		\qquad
        ~ 
        \begin{subfigure}[b]{0.45\textwidth}
                \centering
                \frame{\includegraphics[width=\textwidth]{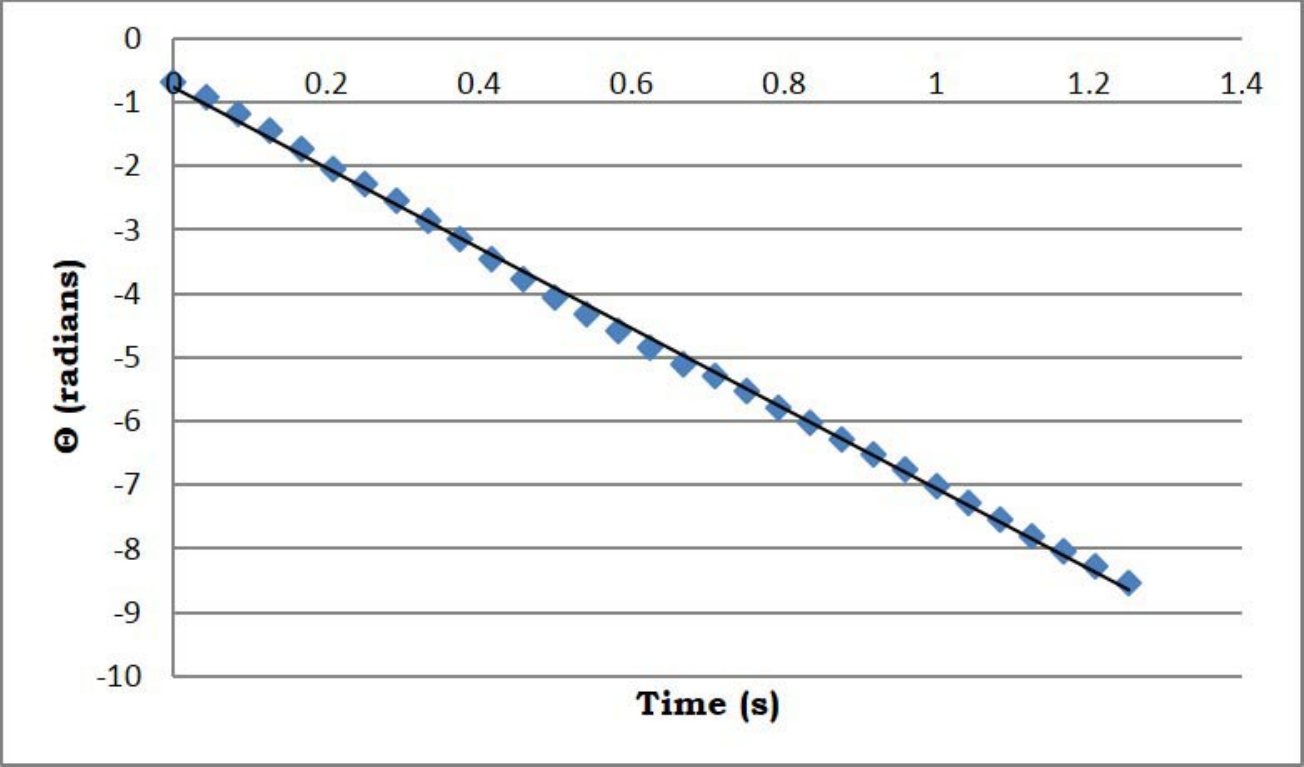}}
                \caption{{\footnotesize Angular position of the mark on the wheel}}
                \label{Fig:WheelAngle}
        \end{subfigure}
        \caption{Video data for the bouncing wheel. The time interval between each two consecutive data points is 1/30 s.}\label{Fig:Wheel}
\end{figure}

In Figures \ref{Fig:WheelVelocity} we show the $v_x$ and $v_y$ of the wheel's center. Based on the vertical velocity, we can clearly see that the wheel underwent three bounces, but the horizontal velocity changed only during the first bounce. The final value of the horizontal velocity is 1.51 m/s.

We can now check explicitly whether (1) is satisfied after the first bounce. Toward that end, we need to determine the final angular speed of the wheel. Fig. \ref{Fig:WheelAngle} displays the angular position of the mark on the rim, with respect to the center. The slope of this graph represents the wheel's angular speed, which comes out to be 6.30 rad/s. 

Measuring the wheel's radius to be 0.246 m, we find that $\omega_{1f} R = 1.55 m/s \approx v_{1xf}$. Thus the horizontal and and angular speeds satisfy (\ref{Smooth}) reasonably well, which confirms the second part of the hypothesis.


\section{The actual coefficient of friction.}\label{Sec:HS_COF}
The attentive reader is now entitled to question the result of Section \ref{Sec:SimpleBounce}. Indeed, it was obtained under the assumption that the force of friction was acting throughout the entire bounce, which has been shown to not be true. Can we fix it and find the actual coefficient of friction? 


In principle, this question can be addressed straightforwardly by using a high-speed camera. If we get a sufficient number of data points during the bounce, we can catch the moment when the force of friction ceases. We can  then apply the same reasoning as in Sec. \ref{Sec:SimpleBounce} to the time interval when both force of friction and normal force are acting. 

We did attempt to ``zoom-into'' the bounce with the help of a 240-fps video camera. Since a typical bounce lasts about 5-10 ms, we only obtained a few data points corresponding to the bounce itself. Besides, the camera we used had a poor resolution in the high-speed mode, which did not allow for accurate position (hence velocity) measuments. What we were able to tell from the data is that - for the typical values of speed - friction vanished in the last quarter of the duration of the bounce. In order to conduct this part of the experiment properly one would need to have at least 1000 fps and a decent resolution at the same time. At the moment, a suitable camera is not a very affordable option for such a basic project.

\section{Discussion}\label{Sec:Discussion}
We have considered a simple experiment that, despite clear expectations, yielded puzzling results and, consequently, made us uncover a deeper layer of physics. This paper may be helpful for physics teachers who tend to (or are planning to) incorporate practical implementation of scientific cycle in their classes, e.g. those using the ISLE approach \cite{ISLE}. We would like to point out, however, that this should probably not be the first such a demonstration, as this paper is an example when``things go wrong''. It should rather be preceeded by a few demos/projects when ``things go right''.

The typical scientific cycle, proceeds as follows
\[
{\rm observation \,\, experiment} \quad \rightarrow \quad {\rm hypothesis, \,\,prediction} \quad \rightarrow \quad {\rm testing \,\,experiment} \hookleftarrow,
\]  
and once the observers are convinced that the hypothesis is valid, they can run {\em application} experiments. In the current paper, the multi-bounce experiment in Sec. \ref{Sec:MultiBounce} was originally meant to be an application experiment: it was supposed to check the {\em value} of the coefficient of friction. Suprisingly it showed that the whole model, which was used successfully in Sec. \ref{Sec:SimpleBounce}, had to be reconsidered.

When doing these experiments, the following measures helped us to achieve a better data accuracy. Due to the angular distortion, a pixel in the middle of a frame is not of the same physical length as one on the rim. Thus it  is best to put the camera as far away from the plane of the ball's trajectory as possible. Along the same lines, angular measurements are typically more accurate than the linear ones. For example, when obtaining the $\theta(t)$ graph in the wheel experiment, only ratios of $x$- and $y$-measurements appear. Therefore the relevant region of the pixel size variation is comparable to the size of the the wheel rather than that of the whole frame.

We did not attempt analyzing experimental uncertainties, but it can certainly be included in the project. One can estimate the uncertainties in the time and position measurements and deduce the error bars for the velocity graphs. Reversing this procedure, one can require a specific accuracy in the velocity data, needed to decide when the force of friction turns off, and figure out the specs of the video camera that would allow to ``zoom into the bounce''. 

We would like to conclude with the following remark. Participating in projects which have hidden assumptions that may not always be true can become invaluable for students. It conveys  a clear message that science is not about getting the right answer on the first attempt by knowing the right recipe. Often times the progress is rather non-linear and making mistakes is quite natural. Not only does one need to know and compute, but also to analyze, evaluate, reflect, improve and eventually arrive at a reasonable answer. Paraphrasing Feynman, ``...in physics, we enjoy the process!''.


\begin{thebibliography}{10}
\bibitem{Logger}
We used Vernier's Logger Pro 

\bibitem{ISLE}
E. Etkina and A. Van Heuvelen, ``Investigative Science Learning Environment – A Science Process Approach to Learning Physics'', in Research-Based Reform of University Physics, edited by E. F. Redish and P. Cooney (AAPT, College Park, MD, 2007); online at per-central.org/per\_reviews/media/volume1/ ISLE-2007.pdf

\end{thebibliography}
\end{document}